\documentclass[reprint,superscriptaddress,amsmath,amssymb,aps]{revtex4-2}

\usepackage[dvipsnames]{xcolor}
\usepackage{float}
\usepackage{graphicx}
\usepackage{dcolumn}
\usepackage{bm}

\newcommand{\vort}{\boldsymbol{\omega}}
\newcommand{\vel}{\mathbf{u}}
\newcommand{\vpeak}{v_{peak}}

\begin{document}

\title{Cosmic Filament Spin from Dark Matter Vortices}

\author{Stephon Alexander}
\affiliation{Department of Physics, Brown University, 182 Hope Street, Providence, RI, 02903, USA}
\affiliation{Center for Computational Astrophysics, Flatiron Institute, New York, NY 10003, USA}
\author{Christian Capanelli}
\affiliation{Department of Physics, Brown University, 182 Hope Street, Providence, RI, 02903, USA}%
\author{Elisa G. M. Ferreira}%
\affiliation{Max Planck Institute for Astrophysics, Karl-Schwarzschild-Str. 1, 85748 Garching, Germany}
\affiliation{Instituto de F\'isica, Universidade de S\~ao Paulo - C.P. 66318, CEP: 05315-970, S\~ao Paulo, Brazil}
\author{Evan McDonough}
\affiliation{Department of Physics, University of Winnipeg, Winnipeg, MB R3B 2E9 Canada}

\begin{abstract}
The recent observational evidence for cosmic filament spin on megaparsec scales (Wang et al, Nature Astronomy 5, 839–845 (2021)) demands an explanation in the physics of dark matter. Conventional collisionless cold particle dark matter is conjectured to generate cosmic filament spin through tidal torquing, but this explanation requires extrapolating from the quasi-linear regime to the non-linear regime. Meanwhile no alternative explanation exists in the context of ultra-light (e.g., axion) dark matter, and indeed these models would naively predict zero spin for cosmic filaments. In this Letter we study cosmic filament spin in theories of ultra-light dark matter, such as ultra-light axions, and bosonic and fermionic condensates, such as superfluids and superconductors. These models are distinguished from conventional particle dark matter models by the possibility of dark matter {\it vortices}.  We take a model agnostic approach, and demonstrate that a collection of dark vortices can explain the data reported in Wang et al.~Modeling a collection of vortices with a simple two-parameter analytic model, corresponding to an averaging of the velocity field, we find an excellent fit to the data. We perform a Markov Chain Monte Carlo analysis and find constraints on the number of vortices, the dark matter mass, and the radius of the inner core region where the vortices are distributed, in order for ultra-light dark matter to explain spinning cosmic filaments.
\end{abstract}

\maketitle

\section{\label{sec:intro}Introduction}

Recent observational evidence (Wang et al.~\cite{Wang_2021}) suggests that some cosmic filaments are spinning. By comparing the redshift and blueshift of galaxies in thousands of filaments, Wang et al.~\cite{Wang_2021} determined that galaxies have velocities perpendicular to the filament axis, consistent with vorticle motions. Meanwhile, it is difficult to theoretically explain the acquisition of angular momentum on megaparsec scales. Vorticity is not easily seeded by density perturbations of a perfect fluid \cite{Christopherson_2011}, and any primordial vorticity is expected to be redshifted away \cite{Lu_2009}. One could try to extend arguments like the tidal-torquing theory introduced in the context of galaxy formation \cite{Peebles_1969, White_1984, Barnes_1987}, but these describe the (quasi) linear regime, and not the non-linear regime needed to describe the filaments of the cosmic web. Recent numerical findings from N-body simulations \cite{Laigle_2014, Codis_2015, Xia_2021,Sheng:2021abt} suggest that standard collisionless cold particle dark matter can produce spinning cosmic filaments, but it has not been demonstrated that this can produce enough spin to explain observations \cite{Wang_2021}, and a detailed analytic understanding is still lacking.

Less effort has been made to understand large-scale rotation in ultra-light dark matter scenarios, such as ultra-light axion or fuzzy dark matter \cite{Hu:2000ke,Hui:2016ltb}, and condensates, both bosonic \cite{Berezhiani_2015,Ferreira:2018wup} and fermionic \cite{Alexander:2018fjp,Alexander:2020wpm,Alexander:2016glq}. In these models (see \cite{Ferreira_2020} for a review), dark matter can have de Broglie wavelengths exceeding the typical inter-particle separation, and is best described by a fluid obeying Euler-like dynamics \cite{Berezhiani_2015,Ferreira_2020,Hui_2021}.   Given their small mass they might reconcile some incompatibilities between the standard cold dark matter and the small scale behaviour of dark matter like the core-cusp problem, missing satellites, and galactic rotation curves \cite{Bullock_2017, Chen_2017, Berezhiani_2018, Niemeyer_2019}. However, as in classical hydrodynamics, one naively expects ultra-light dark matter to be irrotational. Therefore, it seems ultra-light models have little new to offer towards explaining rotations. Yet, vorticity can be introduced in this ultra-light dark matter in the form of defects, namely, vortices. This can be done either by rigid rotation, in analogy to well-understood condensed matter systems \cite{Feynman_1955}, or dynamically through destructive interference (i.e., through density perturbations alone) \cite{Hui_2021}.  This ability for ultra-light dark matter to dynamically acquire vorticity, particularly in the non-linear regime, makes it a compelling candidate to explain Mpc-scale rotations. In particular, for the cosmic filament spin, simulations from \cite{Mocz:2019pyf,Mocz:2019uyd,Mocz:2019pyf,Hui_2021,May:2021wwp} show the presence of interference patterns in the filaments which could in turn support vortices confined to filament-like cylindrical regions. 

\begin{figure}[h!]
    \centering
    \includegraphics[width=0.49\textwidth]{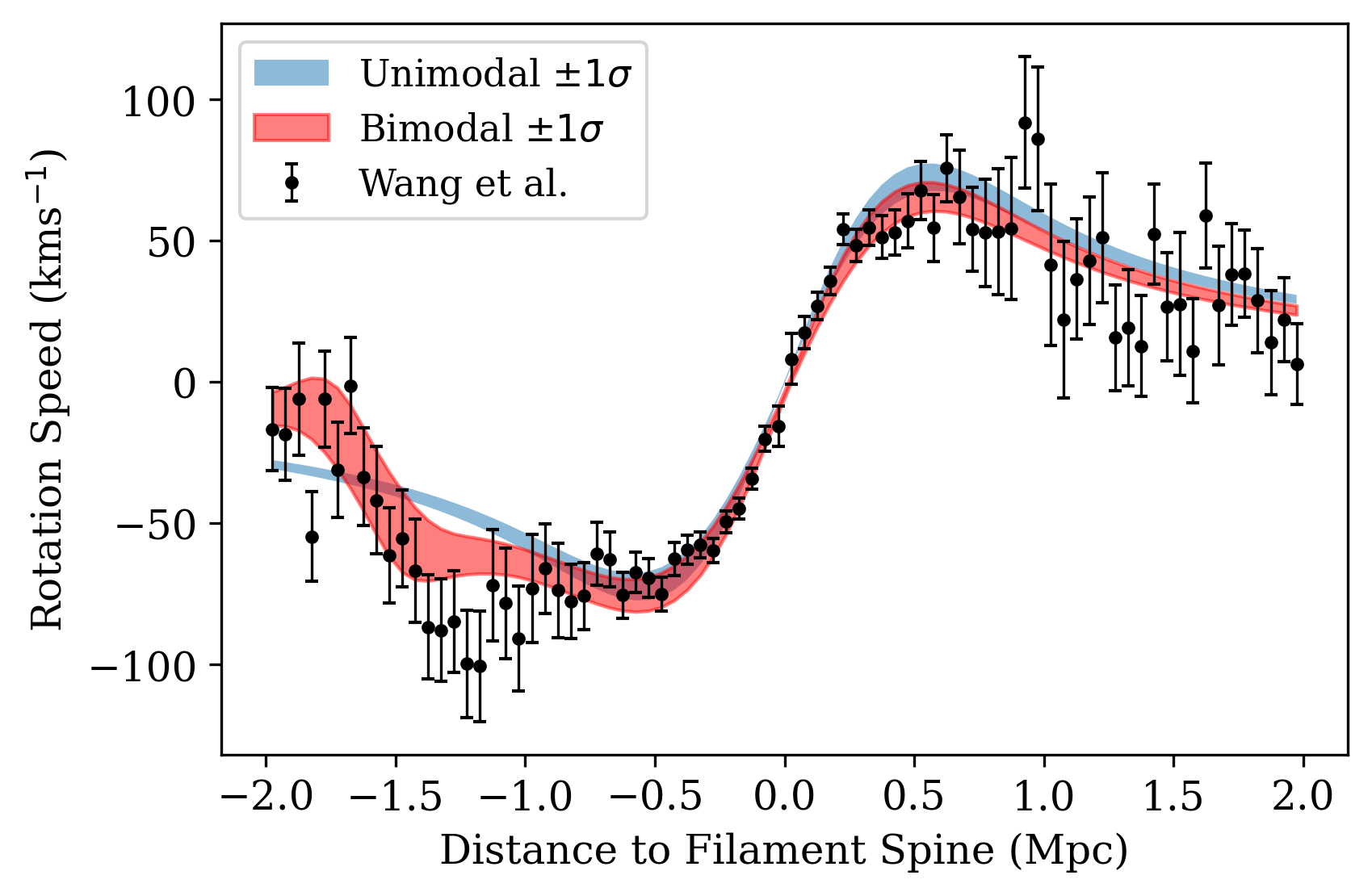}
    \caption{Dark Matter Vortices for Cosmic Filament Spin: Comparison of a Gaussian distribution vortices (blue), a bimodal Gaussian (red), and the data provided by Wang et al.~\cite{Wang_2021}. We plot marginalized $1\sigma$ deviations from the parameter means of an MCMC analysis of the data. }
    \label{fig:bimodal_comparison}
\end{figure}

In this letter we propose dark matter vortices as an explanation for the spin of cosmic filaments. We take a theory agnostic approach to the vortex formation and the underlying particle physics model, and instead focus on the observable signature of vortices. We demonstrate that parallel dark vortices enclosed in a cylindrical volume aligned with the axis of a filament are able to generate rotations at the Mpc scale, and that they can reproduce the behavior seen in \cite{Wang_2021}. Concretely, we find that the data is well fit by a simple Gaussian distribution of vortices about the axis of the filament. We additionally find that a feature present in the data, namely a relative dip in rotation speeds on only one side of the filament,  may be explained by a subdominant population of vortices.  These results are illustrated in Fig.~\ref{fig:bimodal_comparison}.

The structure of this Letter is as follows: in Sec.~\ref{sec:dark_vortices} we review the theory of ultra-light dark matter and dark vortex solutions. In Sec.~\ref{sec:filament-spin} we demonstrate that a distribution of vortices in the inner region of a cosmic filament can generate a net spin, analogous to that reported by \cite{Wang_2021}. In Sec.~\ref{sec:constraints} we perform a Markov Chain Monte Carlo analysis of an ensemble of dark vortices fit to the data reported in \cite{Wang_2021}. We conclude in \ref{sec:discussion} with a discussion of the assumptions made in this work, and directions for future research. We provide additional details in Apps. \ref{app:other_distributions} and \ref{app:vortexvortex}.


\section{Dark Matter Vortices }
\label{sec:dark_vortices}

There are a variety of ultra-light dark matter models in the range of $10^{-24}$eV $\lesssim m\lesssim 1$eV~\citep{Ferreira_2020}. For such low masses, the de Broglie wavelength is in the parsec to kiloparsec range, and this dark matter candidate behaves as wave on galactic scales. We can describe these models by a scalar field $\psi$ which obeys the Gross-Pitaevskii (GP) equation coupled to the Poisson equation \cite{Hui_2021}:
\begin{equation}\label{eq:GP_equation}
    \begin{aligned}
    & i\hbar \dot{\psi} = -\frac{\hbar^2}{2m}\nabla^2 \psi + m\Phi\psi - \frac{g}{m^2} \,|\psi|^2 \\
    & \nabla^2 \Phi = 4\pi G m |\psi(\mathbf{r},t)|^2,
    \end{aligned}
\end{equation}
where $\Phi$ is the gravitational potential, $g$ is the self-interaction coupling, and with the identification that $\rho = m |\psi|^2$. The field can be decomposed as
\begin{equation}\label{eq:wf}
    \psi(\mathbf{r},t) = \sqrt{\frac{\rho(\mathbf{r},t)}{m}}e^{i \Theta (\mathbf{r},t)}.
\end{equation}
The gradient of the phase of the field $\Theta$ encodes the fluid velocity,
\begin{equation}\label{eq:velocity}
    \vel = \frac{\hbar}{m}\nabla \Theta . 
\end{equation}
These variables correspond to the hydrodynamics ones at long distances. Combining these, the Schr\"{o}dinger-Poisson equations can be re-written as hydrodynamical equations, the Madelung equations,
\begin{equation}\label{eq:madelung}
\begin{aligned}
    &\partial_t \rho + \nabla \cdot (\rho \vel) = 0 \\
    & \partial_t \vel + (\vel \cdot \nabla) \vel = -\nabla \Phi - \frac{\nabla P_{\mathrm{int}}}{\rho} + \frac{\hbar^2}{2m^2}\nabla \left( \frac{\nabla^2 \sqrt{\rho}}{\sqrt{\rho}}  \right).
\end{aligned}
\end{equation}
The last term in the second Euler-like equation is only present in this class of models and it is called ``quantum pressure". In the presence of interaction we also have a polytropic type pressure $P_{\mathrm{int}} = (g / 2m^2) \, \rho^2$. In the absence of interaction, these equations describe the fuzzy dark matter model (FDM) (also called wave dark matter), and when interactions are present we have the self-interacting fuzzy dark matter model (SIFDM) \cite{Ferreira_2020}.

As in classical hydrodynamics, one may define a \textit{vorticity}, as
\begin{equation}\label{eq:vorticity}
    \vort \equiv \nabla \times \vel,
\end{equation}
and the \textit{circulation} as,
\begin{equation}\label{eq:circulation}
    \Gamma \equiv \oint_{\partial A} \vel\cdot \mathrm{d}\mathbf{l} = \int_A (\nabla \times \vel)\cdot \mathbf{n} \, \mathrm{da} ,
\end{equation}
where $A$ is a chosen surface in the fluid and $\mathbf{n}$ is the unit normal. Given that the fluid velocity,  Eq.~\eqref{eq:velocity}, is the gradient of a scalar, one immediately infers that $\nabla \times {\bf u}=0$, and hence the vorticity vanishes in ultra-light dark matter scenarios. However, there is a loophole to this: the phase $\Theta$ is undefined when the density $\rho$ vanishes. This allows the vorticity to be finite in highly localized regions, referred to as {\it vortices}.

The field must remain single-valued in the neighborhood of one of these vortices, but it may pick up an extra phase factor so long as it is an integer multiple of $2\pi$. This gives the quantization condition:
\begin{equation}\label{eq:quantization}
    \Gamma = \frac{\hbar}{m}\Delta \Theta = \frac{2\pi n \hbar}{m},
\end{equation}
where $n$ is the \textit{winding number}, so that each vortex carries an integer unit of circulation. The net circulation is then a function of the total winding number contained in the chosen contour. In this way, large circulations can either be achieved with a single vortex or many vortices with small winding number.

Dark Matter vortices can form through a transfer of angular momentum to the dark matter halo when a condensate is formed in its interior, see e.g. \cite{Rindler_Daller_2012,Schobesberger:2021ghi}, or dynamically in regions where the density vanishes, e.g., due to wave interference \cite{Hui_2021}. The first type only happens when we have a superfluid and are expected in the regions where there is a condensate. The second type of vortices are expected to happen in all models of ultra-light dark matter (see \cite{Ferreira_2020} for a description and classification of these models), since interference patterns are expected in all of these models that have this wave like behaviour on small scales.

It is unclear whether vortices can be generated via the transfer of angular momentum  within a filament, given the requirement of condensation. But, from simulations \cite{Mocz:2019pyf,Mocz:2019uyd,Mocz:2019pyf,Hui_2021,May:2021wwp}, we can see that the interference patterns appear in the filaments. One can thus expect that vortices can be formed along these filaments. Although there is considerable theory uncertainty as to the size and abundance of vortices that should be expected, all ultra-light dark matter models are expected to present these interference pattern.  Therefore, in what follows, we remain agnostic to the underlying theory, as well as the formation mechanism, and instead simply consider the observables of vortices. We return to theory expectation for the number of vortices in the Discussion section.

\begin{figure*}[ht!]
    \centering
    \includegraphics[height=4cm]{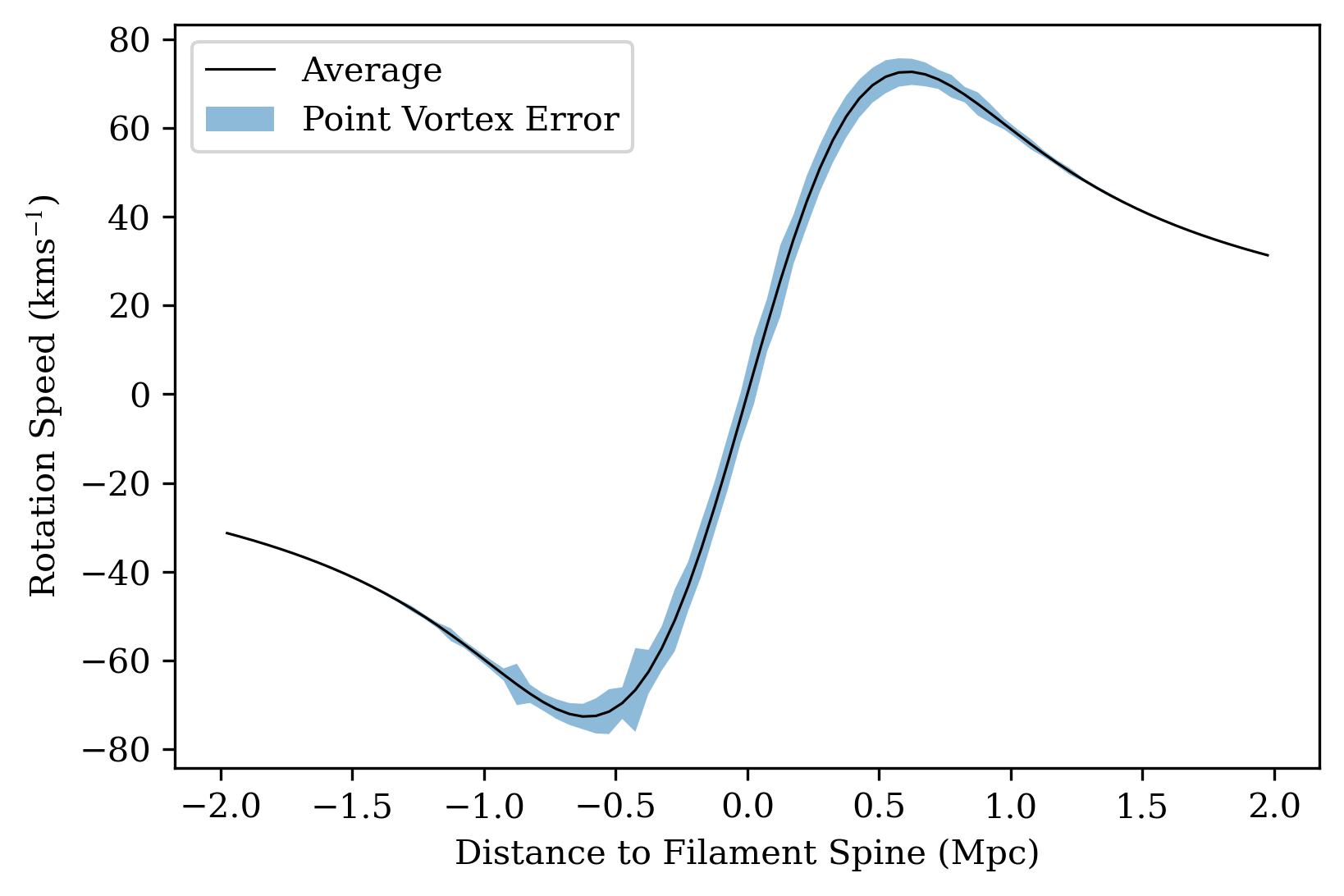}
    \hspace{1cm}
    \includegraphics[height=4cm]{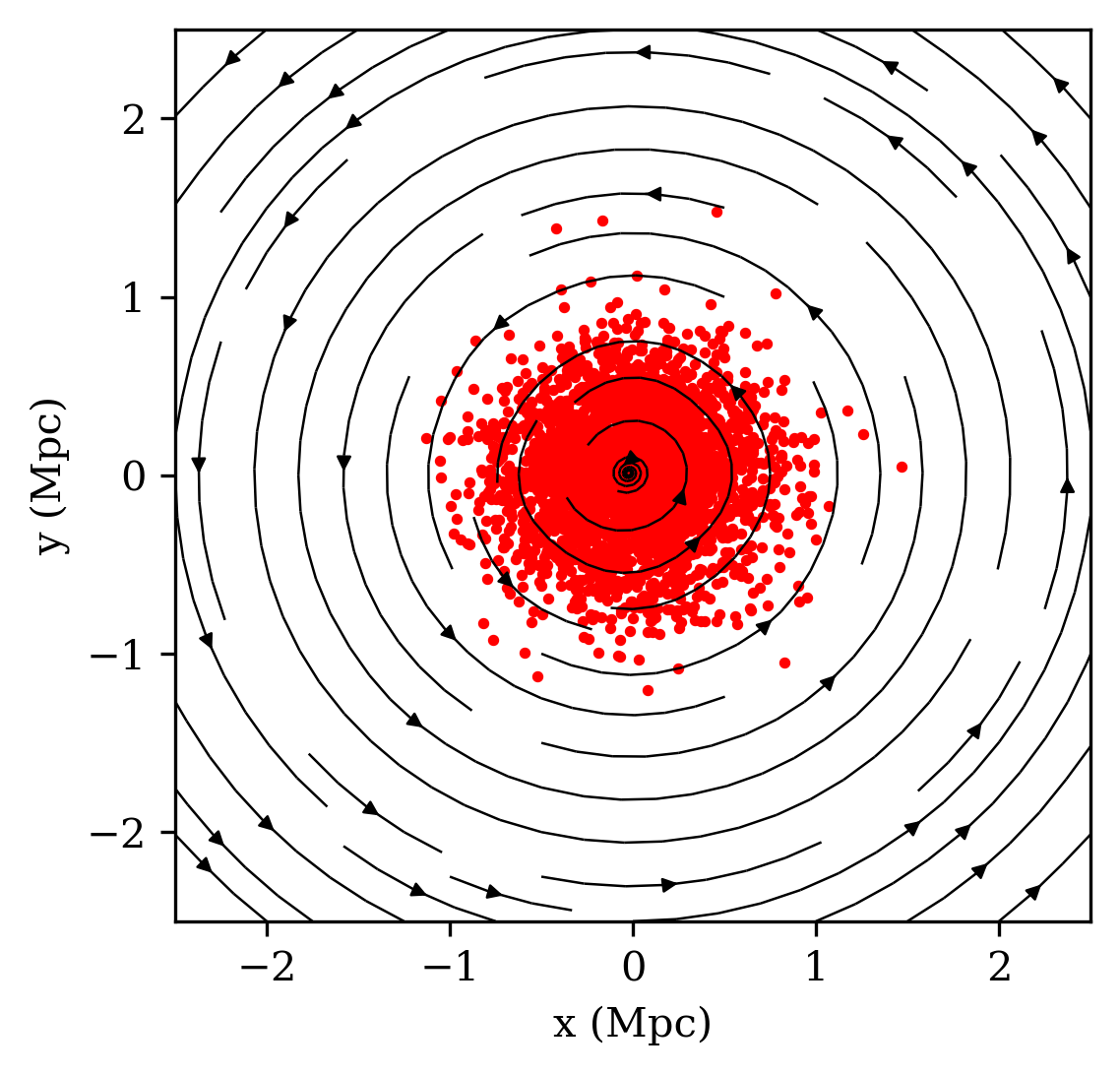}
    \caption{ {\it Left Panel} Fluid rotation curve induced by a collection of vortices. The analytical average is compared to the $1\sigma$ error band that comes from randomly distributing point vortices. Here, $N_{V} =3000 $, $m=10^{-22} \textrm{eV}$, and $R = 0.5 \ \textrm{Mpc}$. {\it Right Panel}: Flowlines of the velocity field induced by $3000$ vortices looking along the filament spine. Red dots indicate vortex lines. Note that the dots are not to scale; as we demonstrate in the App. \ref{app:vortexvortex}, in our analysis the vortex-vortex separation is always greater than the vortex radius.}
    \label{fig:net_flow}
\end{figure*}

\section{Dark Matter Vortices for Cosmic Filament Spin}
\label{sec:filament-spin}

A single vortex carries a unit of angular momentum determined by the winding number. Correspondingly, a collection of dark vortices can generate a net rotation on cosmological (megaparsec) scales.

Consider parallel vortex filaments enclosed in a cylindrical volume that are aligned with the axis of the cosmic filament. We take their cores to have vanishing thickness compared to the radius of the entire filament. In a completely straight vortex there is no self-induction, and thus the system is equivalent to the problem of point vortices in two dimensions \cite{Newton_2011}. It follows that the vorticity, oriented along the axis of the filament, can be decomposed as,
\begin{equation}\label{eq:decomposition}
    \omega = \sum_{i=1}^{N_V} \frac{\Gamma_i}{2\pi}\delta^{(2)}  (\mathbf{x}-\mathbf{x}_i),
\end{equation}
where the sum is over each of $N_V$ vortices, where each vortex has circulation $\Gamma_i$. We take each vortex to have a single quantum of circulation, letting $n=1$ in Eq.~(\ref{eq:quantization}). The corresponding velocity field is
\begin{equation}\label{eq:2D_velocity}
    \vel = \sum_{i=1}^{N_V} \frac{\Gamma_i}{2\pi}\frac{(-(y-y_i),(x-x_i))}{|\mathbf{x}-\mathbf{x}_i|^2},
\end{equation}
where we see that each vortex generates a velocity field with a $1/r$ fall-off. However, the combined effect of all the vortices is more interesting. 

We note that, as with the total number of vortices, there is considerable theoretical uncertainty on the distribution of dark matter vortices within the halo or filaments. Detailed predictions will require full numerical studies of structure formation in these models which are underway by many separate groups, but are very challenging. For simplicity, in this work we assume a Gaussian distribution of vortices in the plane normal to the filament axis, centered at the filament axis. We discuss alternative choices of distributions in App. \ref{app:other_distributions}. We will denote as $N_V$ the number of vortices in a cross-section of the filament; in the simple case of line vortices that span the entirety of the filament, $N_V$ is simply  the total number of vortices.

We now return to the velocity field. From the right panel of Fig.~\ref{fig:net_flow}, one may appreciate that the {\it net} velocity field from many vortices is nearly tangential. One may also notice that, although individual vortices carry a $1/r$ velocity field, their combination recovers a linear growth within a core region. This resembles behavior seen in viscous vortex solutions (cf. \cite{rankine_1895, lamb_1924, kaufmann_1962}) despite the condensate itself being inviscid.

For the purposes of comparing to data, the velocity field of a collection of vortices can be approximated via the analytical average,
\begin{equation}\label{eq:model}
    u_{\theta} = \frac{\Gamma}{2\pi r} \left( 1 - e^{- \frac{r^2}{R^2}}\right),
\end{equation}
with $R$ acting as the effective radius of a composite vortex, and $\Gamma = \sum_i \Gamma_i$ the total circulation. This describes a `typical' realization of vortices with positions drawn from the the Gaussian distribution. In Fig.~\ref{fig:net_flow} we show this analytic average along with the variance from many explicit realizations, from which one may appreciate that the average accurately captures the rotation speed, up to a small error.


\section{Constraints from Data}
\label{sec:constraints}

We perform a Markov Chain Monte Carlo analysis of a collection of vortices fit to the data presented in \cite{Wang_2021}, namely measurements of galactic rotation speed distributed across a distance $\pm 2 $ Mpc to the filament spine. We model the collection of dark matter vortices via the simple model Eq.~(\ref{eq:model}), which describes a symmetric distribution of $N_V$ vortices in a central region $R$, or, equivalently, one composite vortex of radius $R$ that spans the length of the filament. This model has only two free parameters: $R$ and the ratio $N_{V}/m$, where $N_V$ is the number of vortices and $m$ is the dark matter particle mass. We assume uniform priors $R=[0,1]$ Mpc and $N_V/m = [3.5\times 10^{19}, 7\times 10^{86}] \ \textrm{eV}^{-1}$. We use the Python package \texttt{emcee} \cite{emcee_2013} to perform the analysis, and we test convergence of the chains by computing the autocorrelation time $\tau$ of each chain and ensuring that the length of the chains $N_{\rm samples}$ is bigger than $10^{3} \, \tau$, as suggested by the \texttt{emcee} documentation.

The model fit for a Gaussian distribution of vortices is shown in blue in Fig.~\ref{fig:bimodal_comparison}, with posterior distributions given in App.~\ref{app:other_distributions}. We find marginalized parameter constraints,
\begin{eqnarray}
    R && =0.51_{-0.02}^{+0.02}  \ \textrm{Mpc} \\
    \frac{N_{V}}{ m} &&= 2.9^{+0.2}_{-0.2} \times 10^{25} \ \textrm{eV}^{-1} .
\end{eqnarray}
while the best-fit (maximum likelihood) parameters are given by,
\begin{eqnarray}
    R_{ML} && = 0.507\ \textrm{Mpc} \\
    \frac{N_{V}}{ m} \bigg\rvert_{ML}&&= 2.92 \times 10^{25} \ \textrm{eV}^{-1} .
\end{eqnarray}
This fit results in a $\chi^2 = 113.3$, which equates to a reduced $\chi^2$ of $\chi^2/\nu = 1.45$, suggesting a good fit to the data. We also compute the AIC score (BIC), and find AIC (BIC) $=117.4(122.1)$. The contour plot showing the constraints in these parameters can be seen in App.~\ref{app:contour_plots}.

From Fig.~\ref{fig:bimodal_comparison}, blue curve, one may appreciate that the $1\sigma$ uncertainty on the marginalized theory curve is significantly smaller than the error on the data points given in \cite{Wang_2021}. This is driven by the small error bars in the central region of the filament, which is the only region where the error bars fall below $10\%$. For the simple Gaussian distribution of vortices, the central data points effectively fix the distribution and number of vortices, which then fixes the the predictions in the tail regions. 

This simple model is symmetric by construction, however the data is slightly asymmetric, exhibiting a dip in rotation speed around $-1.5 \ \textrm{Mpc}$ (see Fig.~\ref{fig:bimodal_comparison}). This asymmetry is clearly seen in the residual of the fit of our simple symmetric model, showing that a symmetric distribution cannot capture this asymmetry in the data. This asymmetry can be attributed to a deviation of the distribution of vortices from a single Gaussian centred along the filament axis. To demonstrate this, we generalize our model to include a second Gaussian core of vortices (see App.~\ref{app:other_distributions} for details). From Fig. (\ref{fig:bimodal_comparison}),  orange curve, one may appreciate that the bimodal distribution provides an excellent fit to the data, and indeed we find $\chi^2=76.28$ ($\chi^2/\nu = 1.02$) and AIC(BIC) $=86.28(98.19)$, indicating a substantial improvement on the fit.

As a qualitative check on these results, we can compare the best-fit parameters with order of magnitude estimates using Eq.~(\ref{eq:circulation}). If the velocity field of halos around the filament really is tangential, then the total circulation is $\Gamma = 2\pi \vpeak R$. To get the ratio $N_v / m$, we then count how many vortices can be supported by this total circulation. Rearranging Eq. (\ref{eq:quantization}) we get
\begin{equation}\label{eq:ratio}
    \frac{N_{V}}{m} = \frac{\Gamma}{2\pi \hbar} = \frac{\vpeak R}{\hbar},
\end{equation}
which, for $\vpeak \sim 70  \textrm{km s}^{-1}$ and $R \sim \textrm{Mpc}$, gives $N_V/m \sim 10^{25} \ \textrm{eV}^{-1}$, in agreement with the maximum likelihood parameters.


\section{Discussion}\label{sec:discussion}

The model presented here demonstrates that dark matter vortices can account for the spin of cosmic filaments \cite{Wang_2021}. These vortices can be arranged very simply while still recovering the observed rotation curves. This provides a competing explanation to tidal torquing of cold particle dark matter as the origin of cosmic filament spin. 

From the results obtained in Section~\ref{sec:constraints}, namely given the value of the $N_V / m$ obtained, we can see that an ultra-light dark matter candidate with a wide range of masses can explain the data. For example, a mass of order $10^{-22} \, \mathrm{eV}$ and with roughly $3000$ vortices spanning the filament can explain the data. This is well within the regime where the average approximation Eq.~\eqref{eq:model} is valid, providing an a posteriori justification for this simplifying assumption.

The configuration found here in the fits of the rotation of the filament is expected to be produced in both of the vortex formation mechanisms possible for these ULDM models. For the vortices formed in the regions where destructive interference occurs, it is expected that we have 3 vortices per de Broglie area \cite{Hui_2021}. Given this, considering a FDM particle with mass $m = 10^{-22} \, \mathrm{eV}$ that has a Broglie wavelength  order of kpc, we could have from $\mathcal{O}(10^3) - \mathcal{O}(10^5)$ vortices, depending on their spacing and distribution. Therefore, having 3000 vortices in this Mpc filament as shown to be necessary for our fit to explain the net rotation of the filament in \cite{Wang_2021} is easily achievable and expected in this model. For the second mechanism of formation of vortices, i.e., in a superfluid through the transfer of angular momentum, vortices are formed when the angular velocity is bigger than the critical angular velocity, which on the mass and coupling of the particles in the superfluid.  We are still lacking simulations in order to have a realistic estimation of the formation of the vortices in these systems.  Some rough estimations were made in the literature for the case of the dark matter halo. In the case of a weakly coupled superfluid described in Eq. (\ref{eq:GP_equation}), \cite{Silverman:2002qx} predicts 340 vortices in the M31 halo for a dark matter particle mass $m = 10^{-23}$ eV. In the case of the dark matter superfluid \cite{Berezhiani_2015} where the particle has a mass of the order of $m = 1$ eV, $N = 10^{23}$ vortices are expected. Given those estimates, the values found in our fit are realistic.

We have also assumed in this analysis that the vortices can be treated as non-interacting, analogous to the dilute instanton gas approximation used in quantum field theory \cite{Shuryak_1981, Borsanyi_2016}. In realistic condensed matter systems, and their dark matter analogs, there can be vortex-vortex collisions and reconnections, as well as small interaction between vortices \cite{Banik:2013rxa}. However, similar to the averaging assumption, the number of vortices needed to explain the data is well within the regime where these effects can be neglected. We can estimate this as follows: In order for the vortices to be well-separated, within a cross-section of cylinder we require that the area contained within vortices, $N_V \pi r_V^2 $, is much less than the area of the filament in which the vortices are predominantly distributed, $\pi R^2$. For $r_V=\mathcal{O}(\mathrm{kpc})$ (for $m=10^{-22}$ eV) and $R = \mathcal{O}(\mathrm{Mpc})$ this provides a bound $N_V \ll 10^6$, well above the number required ($\approx 3000$ for $m=10^{-22}$  eV) to explain the data.  In the case of a condensate, the vortex size is determined by the healing length of the superfluid, which is in turn determined by the self-interaction strength.  In this case we again find that dark matter vortices can explain the data while remaining well separated (see App.~\ref{app:vortexvortex} for details).

A third assumption simplifying assumption in this work, see Eq.~\eqref{eq:model} and the blue curve in Fig.~\ref{fig:bimodal_comparison}, is that the distribution of vortices is symmetric. In a real physical situation one expects some degree of asymmetry,  e.g., it is easily conceivable that vortices would form kinks or arrange themselves in a more complicated manner.  Indeed a small amount of asymmetry is exhibited by the data (though it remains possible this asymmetry is due to systematic errors). Taking the data at face value, we generalize our simple model to accommodate an asymmetry by modifying in the distribution of vortices to a  bimodal Gaussian distribution, see App.~\ref{app:other_distributions}. From Fig.~\ref{fig:bimodal_comparison}, red curve, one may appreciate that this distribution provides an excellent fit to the data.  We additionally note that the vortices are expected to form following the distribution and shape of the filament, and therefore, if the filament formed itself formed an asymmetric shape or an asymmetric distribution, the vortices can be expected to follow this. This provides another justification to vary the vortex distribution, as done in App.~\ref{app:other_distributions}.

These assumptions aside, in this work we have focused solely on the observational signature of vortices, and not on their formation mechanism. In doing so we remain agnostic as to the model realization. As we have discussed, vortices can be formed both by destructive interference or by angular momentum of condensate dark matter. It is not the goal of this work to identify which of these mechanisms is most likely responsible for the formation of these vortices, but to show that, given a mechanism to generate vortices, said vortices can generate angular momentum on Mpc scales. This is a very generic feature and independent of the formation mechanism.

Finally, we note that cosmic filament spin may be complementary to the strong gravitational lensing signal of dark matter vortices \cite{Alexander:2019puy,Hui_2021}, and of dark matter substructures generally (see, e.g.,~\cite{Alexander:2019qsh}). While filament spin is sensitive to the net vorticity of the vortices, the lensing signal is sensitive only to the total mass contained therein. A further interesting and open question is whether these varying observational probes can distinguish between fermionic superfluids \cite{Alexander:2018fjp,Alexander:2020wpm,Alexander:2016glq} and bosonic superfluids \cite{Berezhiani_2015,Ferreira:2018wup}. We leave these interesting questions to future work.\\

\noindent {\bf Acknowledgements}

The authors thank the authors of \cite{Wang_2021} for providing their data. The authors thank Vyoma Muralidhara and Sherry Suyu for fruitful discussions regarding the MCMC analysis. The authors thank Arthur Kosowsky for helpful and insightful comments.

\appendix

\section{Varying the Vortex Distribution}
\label{app:other_distributions}

Here we consider two additional distributions of vortices, a uniform distribution and bimodal distribution.  

\begin{figure*}[ht]
    \centering
    \includegraphics[width=0.45\textwidth]{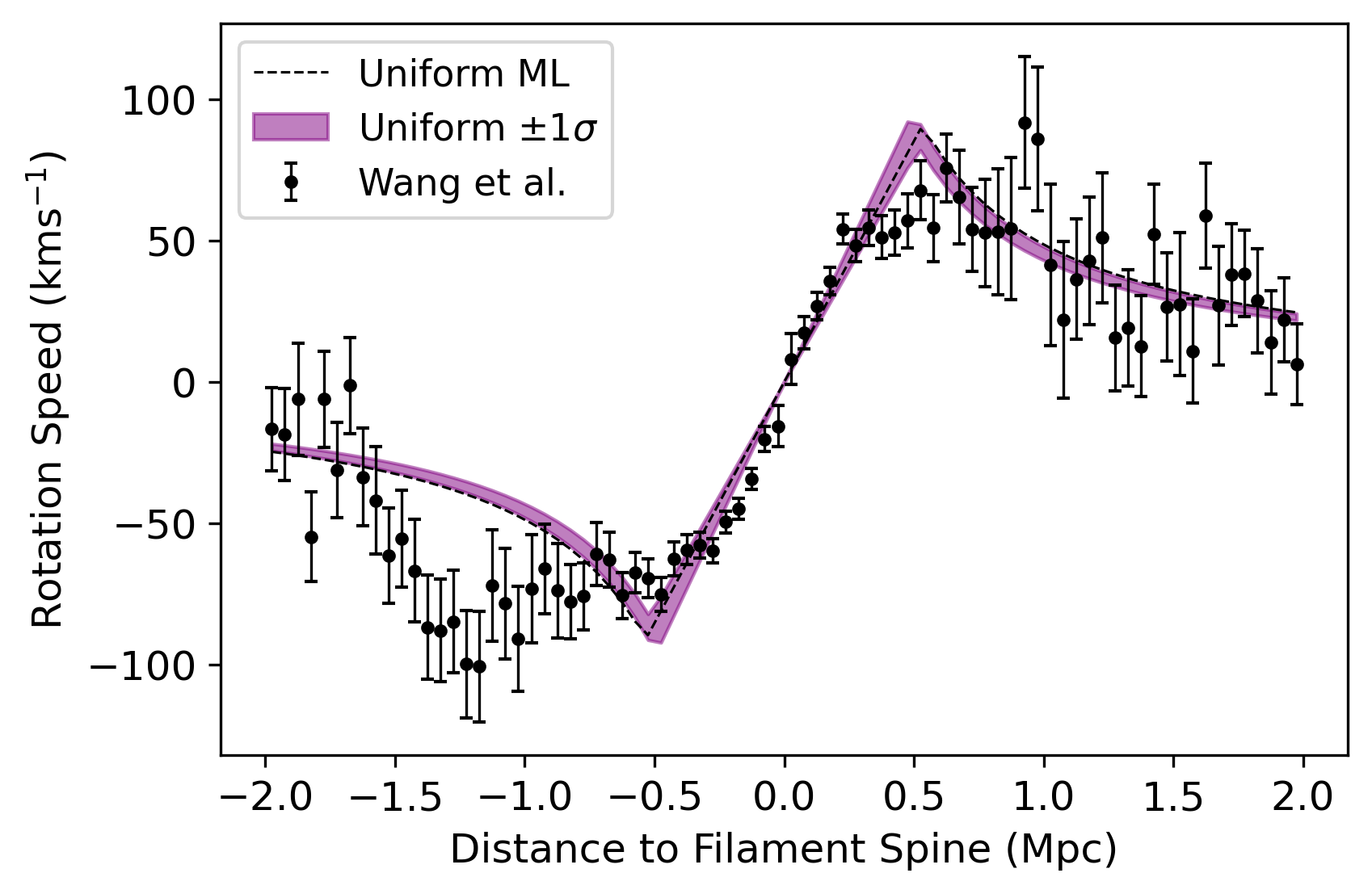}
    \includegraphics[width=0.4\textwidth]{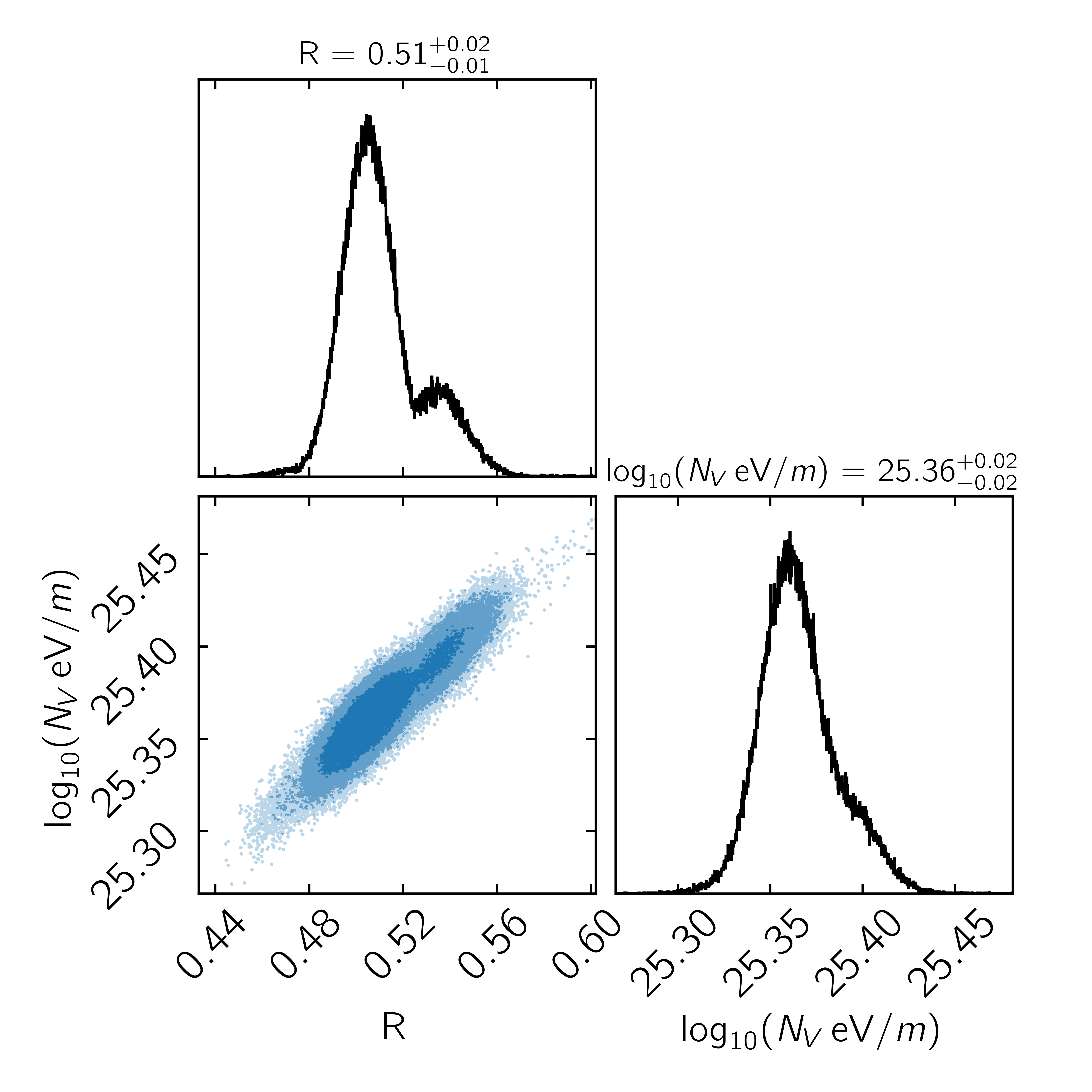}
    \caption{Results and parameters constraints in the fit of a uniform distribution of vortices to the data in \cite{Wang_2021}.}
    \label{fig:model_fit_uniform}
\end{figure*}

First, place vortices uniformly in the circle. If we imagine expanding a contour centered at the origin, then the number of enclosed vortices grows with the area as $r^2$. Outside the core radius, the number of vortices remains constant. Then using Eq.~(\ref{eq:circulation}) one can solve for the velocity~\cite{rankine_1895}:
\begin{equation}\label{eq:Rankine}
u_{\theta} = \frac{\Gamma}{2\pi}
\begin{cases}
    \frac{r}{R} & r \leq R \\
    \frac{1}{r} & r > R
\end{cases}
\end{equation}
Performing the same analysis as done in Section~\ref{sec:constraints}, we find marginalized parameter constrains $R=0.51_{-0.01}^{+0.02} \ \textrm{Mpc}$ and $N_V / m = 2.3^{+0.1}_{-0.1} \times 10^{25} \ \textrm{eV}^{-1}$. The resulting fit is shown in Fig.~\ref{fig:model_fit_uniform}.

It may be desirable to break the axial symmetry of the distribution. A simple way to do this is to introduce a second Gaussian peak. This extra vortex core has parameters $(R_L, \ \delta, \ f  )$ corresponding to its effective radius, offset from the other core, and comparative circulation, respectively. The label ``L" signifies only the convention that the second core is offset left of the original core in the chosen coordinates. The bimodal model is then simply a superposition of the unimodal vortex distribution described by Eq.~(\ref{eq:model}):
\begin{equation}\label{model_bimodal}
    \begin{aligned}
       & u_{\theta} \big(r+\delta, \ R_L, \ f \tfrac{N_V}{m}\big ) + u_{\theta}\big( r , \ R_R, \ \tfrac{N_V}{m}\big)
    \end{aligned}
\end{equation}
The best-fit parameters are listed in Tab.~\ref{tab:bimodal_fit_params}. A comparison of this model to the unimodal Gaussian fit can be seen in Fig.~\ref{fig:bimodal_comparison}. 

\begin{table}[h!]
    \centering
    \begin{tabular}{|c|c|c|c|}
        \hline
         & Uniform & Unimodal & Bimodal \\ 
         \hline
         $ \chi^2$ & 203.4 & 113.4 & 76.28\\
         AIC & 207.9 & 117.4 & 86.28\\ 
         BIC & 212.6 & 122.1 & 98.19\\
         \hline
    \end{tabular}
    \caption{Comparison of different goodness of fit measures between unimodal and bimodal vortex distributions. }
    \label{tab:goodness_of_fit}
\end{table}

\begin{table}[h!]
    \centering
    \begin{tabular}{|c|c|}
      \hline
    Parameter & Constraint \\
    \hline
    $R_L$ (Mpc) &  $0.22 (0.16)_{-0.07}^{+0.09}$ \\
    $R_R$ (Mpc) & $0.50 (0.50 ) \pm 0.02$ \\
    $\delta$ (Mpc) &  $1.61 ( 1.56)_{-0.06}^{+0.10}$ \\
    $f$ &   $-0.13 (-0.11)_{-0.04}^{+0.03} $ \\
    $N_V/m$ (eV$^{-1}$) & $2.7 (2.8)^{+0.2}_{-0.2} \times 10^{25}$\\ 
    \hline
    \end{tabular}
    \caption{The mean (best fit) $\pm 1\sigma$ constraints on a bimodal Gaussian distribution of vortices fit to the spinning cosmic filament data of Wang et al.~\cite{Wang_2021}.}
    \label{tab:bimodal_fit_params}
\end{table}

\section{The Dilute Vortex Gas approximation}
\label{app:vortexvortex}

In order to think of vortices as a dilute gas, we require that the vortices be well separated inside the filament. As a simple condition to enforce this, we require that in a cross-section of the filament, the region where the vortices are predominantly distributed, $\pi R^2$, is much larger than the area contained within the vortices themselves, $N_V \pi r_V^2$, where $r_V$ is the vortex core radius. This amounts to the condition that $N_V \ll (R/r_V)^2$.

The core radius of a singly quantized vortex is of order the healing length $\xi$ \cite{pitaevskii2003bose}. In order to obtain an estimate, we specialize in the case of a superfluid here. The healing length in this case is given by,
\begin{equation}
    \xi = \frac{\hbar}{\sqrt{2\bar{\rho}g}}
\end{equation}
where $\bar{\rho}$ is the ambient dark matter density of the filament and $g$ is the self-interaction strength of the condensate. From this we can derive a lower bound on the coupling $g$ such that the healing length is much less than the typical vortex-vortex separation, thereby justifying the use of the dilute vortex gas approximation.

As a simple benchmark, we consider the lower bound on $g$ in order for vortices to form in a typical halo. Following the results of \cite{Rindler_Daller_2012} we estimate that that $g\gg g_H \sim 10^{-64} \, {\rm eV} {\rm cm}^3 \simeq 10^{-58} \, {\rm Mpc}^2$ for vortices to form. Meanwhile, the recent analysis in \cite{refId0} reports that the density contrast inside of a cosmic filament is $\delta = 19 ^{+27} _{-12} $, corresponding to a density of $\bar{\rho} \simeq 10^{21} \, {\rm eV}^4 \simeq 10^{84} \, {\rm Mpc}^{-4}$. From this we find,
\begin{equation}
    r_V \simeq \xi \simeq \sqrt{ \frac{g_H}{ g}} 10^{-13} \, {\rm Mpc},
\end{equation}
where $g_H \equiv 10^{-64} \, {\rm eV} {\rm cm}^3$ is the threshold coupling for these superfluid vortices to form in a typical halo. Given $R\sim $Mpc, the dilute vortex gas assumption provides a combined bound on the coupling $g$ and the number of vortices $N_V$ as
\begin{equation}
    \frac{g_H}{g} N_V \ll 10^{26}
\end{equation}
For $g$ at the threshold coupling for vortex formation, $g=g_H$, the dilute gas approximation requires $N_V < 10^{26}$. For larger (yet still extremely weak) couplings the number of vortices can be much larger. Comparing to the number of vortices required to explain the data (e.g.,~$N_V \sim 3000$ for $m=10^{-22}$ eV), we conclude that our analysis is well within the dilute gas approximation.

\section{Posterior Distributions}
\label{app:contour_plots}

\begin{figure*}[h]
    \centering
    \includegraphics[height=7cm]{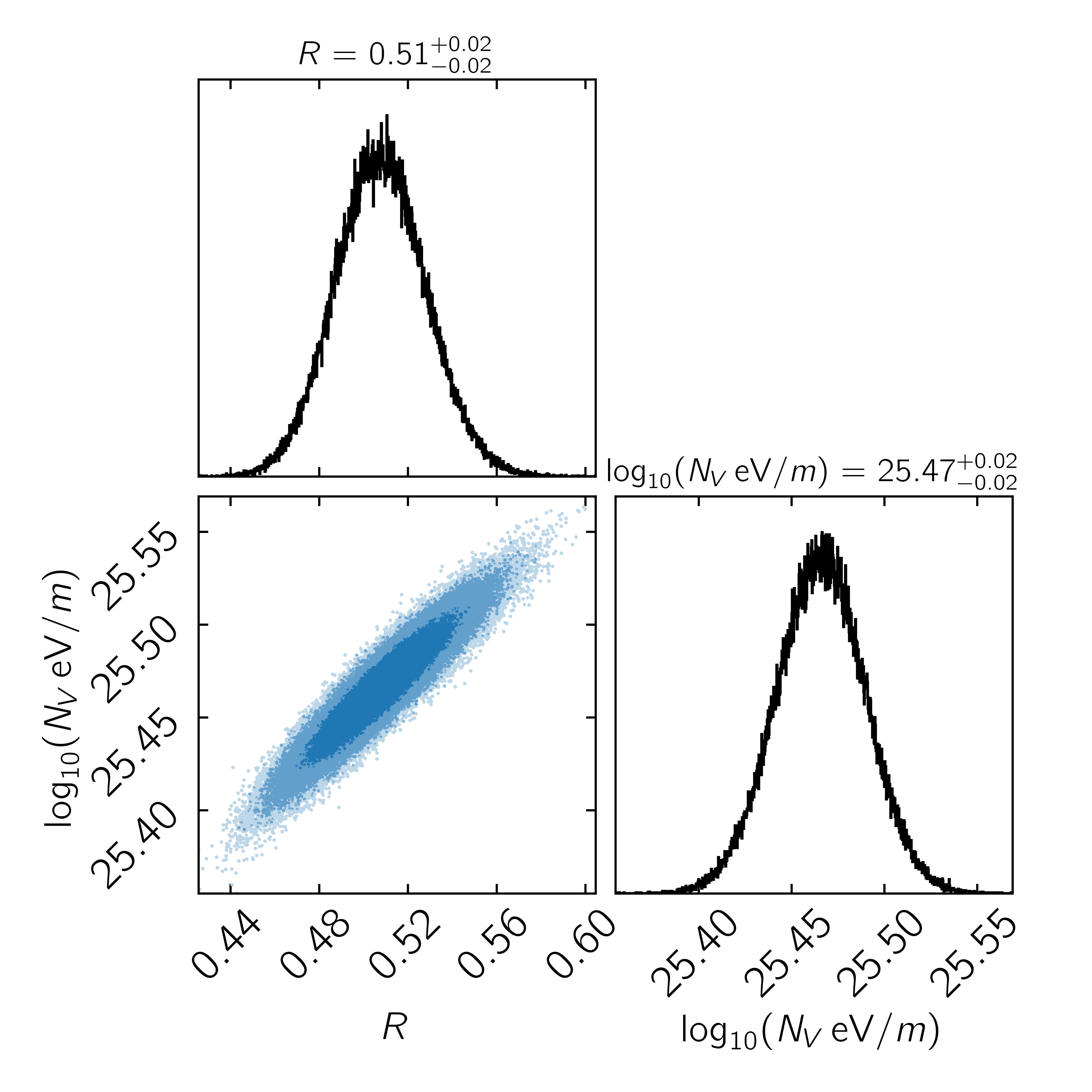}
    \caption{\label{fig:corner_plots} Parameter constraints in the fit of a Gaussian distribution of dark matter vortices to the cosmic filament spin data in \cite{Wang_2021}.}
\end{figure*}

\begin{figure*}[h]
    \centering
    \includegraphics[width=0.7\textwidth]{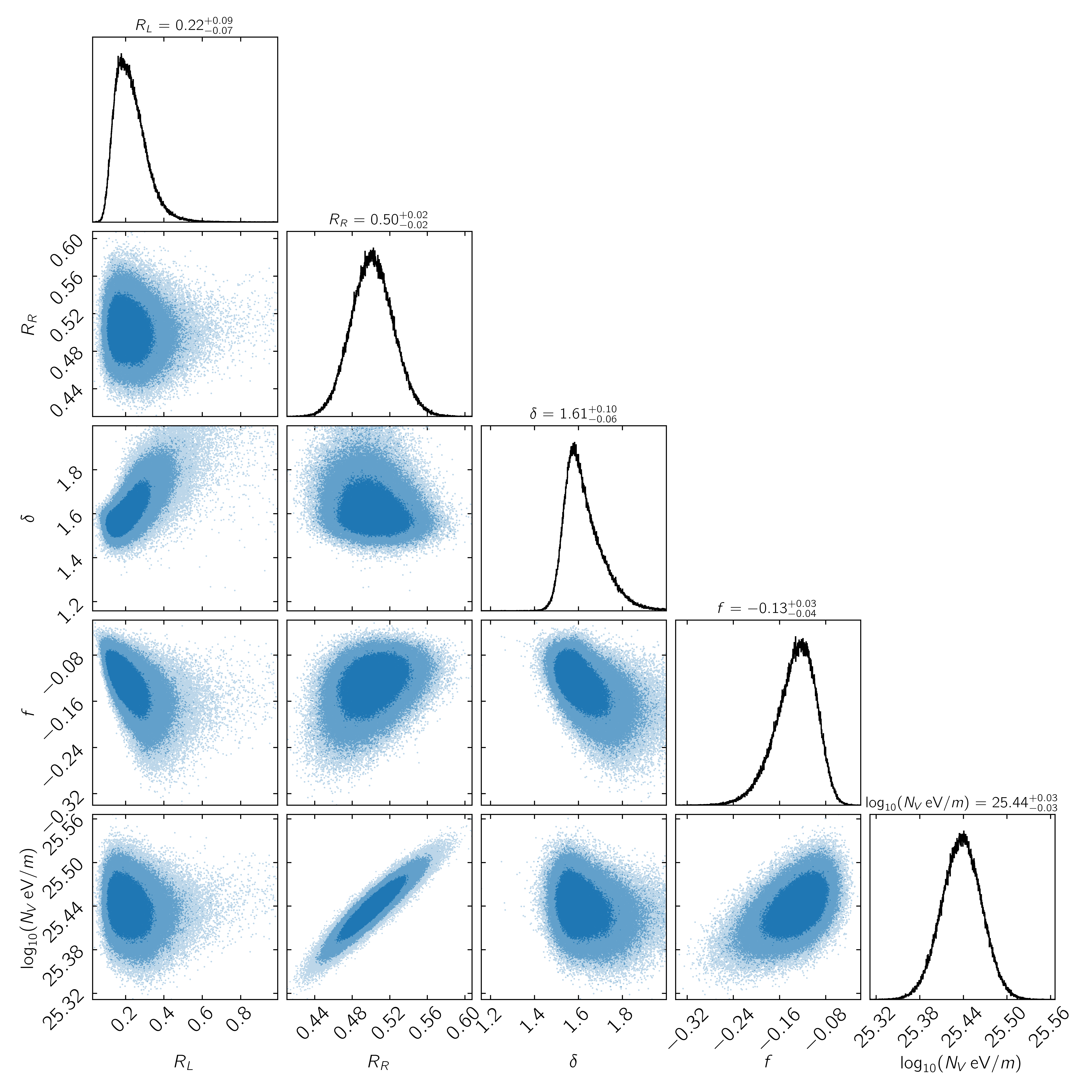}
        \caption{\label{fig:corner_plots_2} Parameter constraints in the fit of a \textit{bimodal} Gaussian distribution of dark matter vortices to the cosmic filament spin data in \cite{Wang_2021}.}
\end{figure*}

The posterior distributions of the fit of a Gaussian distribution of dark matter vortices to cosmic filament spin data is given in Fig.~\ref{fig:corner_plots}, while in Fig.~\ref{fig:corner_plots_2} we show the corresponding distributions in the case of a bimodal Gaussian distribution.

\bibliographystyle{JHEP}
\bibliography{filament}

\end{document}